Structural properties and anisotropic electronic transport in SrIrO$_3$ films

K. R. Kleindienst[1], K. Wolff[1], J. Schubert[2,3], R. Schneider[1], and D. Fuchs[1]

[1]Karlsruher Institut für Technologie, Institut für Festkörperphysik, 76021 Karlsruhe, Germany

[2]Forschungszentrum Jülich, Peter Grünberg Institut (PGI-9), 52425 Jülich, Germany

[3]JARA-FIT - Jülich-Aachen Research Alliance for Fundamentals of Future Information Technology,
Forschungszentrum Jülich GmbH, 52425 Jülich, Germany

ABSTRACT

Perovskite SrIrO$_3$ (SIO) films epitaxially deposited with a thickness of about 60 nm on various substrate materials display nearly strain-relieved state. Films grown on orthorhombic (110) DyScO$_3$ (DSO) are found to display untwinned bulk-like orthorhombic structure. However, film deposition on cubic (001) SrTiO$_3$ induces a twinned growth of SIO. Resistance measurements on the SIO films reveal only weak temperature dependence, where the resistance $R$ increases with decreasing temperature $T$. Hall measurements show dominant electron-like transport throughout the temperature range from 2 K to 300 K. At 2 K, the electron concentration and resistivity for SIO on STO amount to $n_e = 1.4 \times 10^{20}$ cm$^{-3}$ and 1 mΩcm. Interestingly, the film resistance of untwinned SIO on DSO along the [1-10] and the [001] direction differs by up to 25% indicating pronounced anisotropic electronic transport. The anisotropy of the resistance increases with decreasing $T$ and displays a distinct maximum around 86 K. The specific $T$-dependence is similar to that of the structural anisotropy $\sqrt{(a^2+b^2)}/c$ of bulk SIO. Therefore, anisotropic electronic transport in SIO is very likely induced by the orthorhombic distortion. Consequently, for twinned SIO films on STO anisotropy vanishes nearly completely. The experimental results show that structural changes are very likely responsible for the observed anisotropic electronic transport. The strong sensitivity of the electronic transport in SIO films may be explained in terms of the narrow electron-like bands in SIO caused by spin-orbit-coupling and orthorhombic distortion.

I. INTRODUCTION

The strong spin-orbit-coupling (SOC) in 5$d$ iridium-based transition-metal oxides results in comparable energy scales of the electron correlation, electronic bandwidth, and SOC [1,2], which makes these materials promising candidates for the emergence of new topological phenomena or quantum states [3-7]. Systematic dimension-controlled physical properties are observed in the Ruddlesden-Popper series Sr$_{n+1}$Ir$_n$O$_{3n+1}$ ($n = 1$, 2, and ∞) [8]. In Sr$_2$IrO$_4$ ($n = 1$) the crystal-field splitting and SOC lead to new spin-orbital mixed states. The five 5$d$ electrons of Ir$^{4+}$ result in a filled low-energy J$_{eff}$ = 3/2 quartet band and a half-filled high-energy J$_{eff}$ = ½ doublet band. Moderate Coulomb interaction opens a Mott-gap which leads to an antiferromagnetic insulating ground state with $J_{eff} = ½$ [9]. In the perovskite SrIrO$_3$ (SIO) ($n = \infty$), the network of corner-shared IrO$_6$ octahedra provides a better hybridization between Ir 5$d$ orbitals and O 2$p$-orbitals that favors a paramagnetic semi-metallic ground state [10,11]. Tilts and rotations of the relatively rigid IrO$_6$ octahedra, i. e., an in-phase rotation along the $c$-axis and anti-phase rotations along the $a$- and $b$-axes, cause an enlargement of the perovskite unit cell by √2×√2×2 with an, according to the Glazer notation [12], $a^-a^-c^+$ octahedral tilt pattern and an orthorhombic structure with space group *Pbnm* (62) [13].

However, the metastable form of SIO prevents single-crystal growth under ambient pressure, where only the monoclinic modification with C2/2 (15) dominates the ambient phase [14]. Nevertheless, SIO could be successfully synthesized in polycrystalline form under pressure ($p \approx$ 40 kbar) [15] or stabilized by the epitaxial growth of thin films [16-20]. Therefore, epitaxially grown SIO films are of current interest to explore the system. In addition, SIO films might act as a key building block for engineering topological phases at interfaces and in heterostructures [21,22].

Despite the larger coordination and dimensionality compared to the quasi two-dimensional insulating counterpart Sr$_2$IrO$_4$, the electronic bandwidths of SIO are found to be narrower displaying a semi-metallic electronic structure. In particular, its Fermi surface consists of multiple heavy hole- and light electron-like sheets. The 2 – 6 times lighter quasiparticle mass of the electrons allows them to dominate electronic transport explaining the commonly observed electron-like single-type carrier transport in SIO [19, 20,23]. Very recently it was shown that in SIO thin films the electronic structure is controlled by a subtle interplay between octahedral rotations, SOC, and dimensionality [11,24,25]. This paves the way for a distinct tuneability of the physical properties by epitaxial strain and film thickness. For example, angle resolved photoemission spectroscopy (ARPES) and first-principle calculations [11] show that substantial changes in the electronic structure and the physical properties of SIO are achieved by subtle changes in the structure and rotation angles of the IrO$_6$ octahedra. For SIO films on SrTiO$_3$ (STO) substrates bulk-like electronic structure is observed for a film thickness $t > 9$ unit cells (3.2 nm), i. e., paramagnetic metallic behavior with a partially filled $J_{eff} = ½$ band. In contrast, for $t \leq 3$ unit cells a distinct charge gap opens leading to a metal-insulator transition, which on a first glance appears to be in analogy to the metal-insulator transition in the Ruddlesden Popper iridates driven by dimensionality with decreasing $n$ [24,25]. However, the gap-opening is accompanied by a structural transition and thus very likely not caused by the decreased film thickness alone. Specifically, a suppression of in-plane rotations of the IrO$_6$ octahedra is observed which has been discussed in terms of constraints upon octahedral in-plane rotations imposed by the cubic STO substrate.

In this paper, we demonstrate and discuss anisotropic electronic transport of SIO films epitaxially grown on various substrate materials. The relatively thick films ($\approx 60$ nm) display nearly strain-relaxed structural properties. Anisotropic electronic transport was found for untwinned (110) oriented SIO films on (110) DyScO$_3$ (DSO), whereas for twinned SIO on (001) STO anisotropy vanishes nearly completely. The distinct *T*-dependence of the resistance anisotropy indicates that the reason for electronic anisotropy is most likely related to the structural anisotropy along the [1-10] and [001] direction of SIO. The narrow multiple band structure caused by the orthorhombic distortion and SOC in combination with the semi-metallic behavior of SIO may be the reason for the high sensitivity of the electronic transport with respect to structural changes.

## II. EXPERIMENTAL

Epitaxial perovskite SrIrO$_3$ (SIO) films were grown by pulsed laser deposition using home-made single-phase (monoclinic) polycrystalline SrIrO$_3$ targets that were prepared by standard solid-state sintering under ambient pressure. The laser ablation results in a sizeable non-stoichiometric redeposit of ablated materials on the target surface. For this reason, targets were re-polished before each film deposition. The films were deposited at a substrate temperature $T_s = 700$°C. Before deposition, the substrates were annealed in vacuum at 700°C for about half an hour. The thermodynamic stability of neighbored phases of the Ruddlesden-Popper series, i. e., Sr$_2$IrO$_4$ and Sr$_3$Ir$_2$O$_7$ and the large volatility of Ir-oxide species generate a large sensitivity of the Sr/Ir ratio in the film with respect to the deposition parameters, even though single phase targets of SrIrO$_3$ are used [26,27]. Film composition was checked by Rutherford backscattering spectrometry (RBS) using He$^+$ ions with an energy of 1.4 MeV. The computer software RUMP provided the numerical analysis of the RBS-data [28].

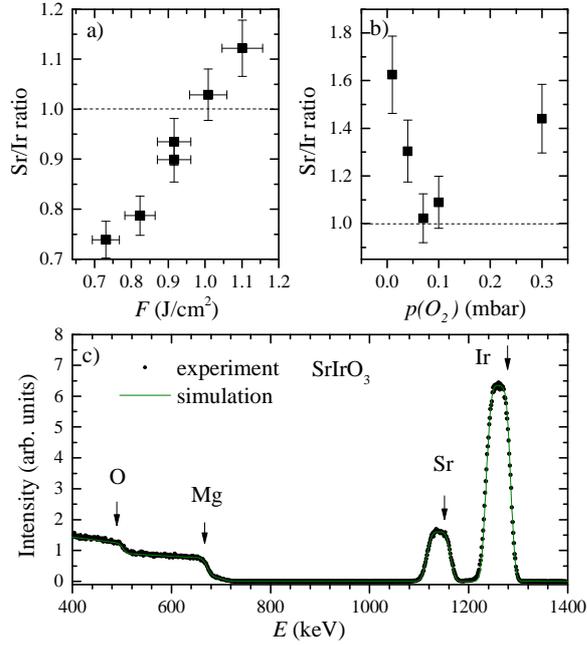

**FIG. 1.** (a) Sr/Ir ratio of deposited films versus laser fluence $F$ for $p(O_2) = 0.1$ mbar and (b) versus $p(O_2)$ for $F = 1$ J/cm$^2$. The Sr/Ir ratio was determined by Rutherford backscattering spectrometry (RBS). Error bars are calculated from standard deviations by measuring multiple samples of each type. Dashed lines indicate stoichiometric composition. (c) RBS spectrum (full circles) for film deposited at $p(O_2) = 0.1$ mbar and $F = 1$ J/cm$^2$ on a MgO substrate . The chemical elements corresponding to the peaks are indicated. A fit to the spectrum is shown by the solid line that reveals stoichiometric composition of SrIrO$_3$.

A linear dependence of the Sr/Ir ratio was found for laser fluence $0.7$ J/cm$^2$ $\leq F \leq 1.1$ J/cm$^2$ and an oxygen partial pressure $0.01$ mbar $\leq p(O_2) \leq 0.1$ mbar, see Fig. 1. This in turn allows a precise control of the chemical composition. Stoichiometric conditions are obtained for $p(O_2) \approx 0.1$ mbar and $F \approx 1$ J/cm$^2$. The deposition rate amounts to about 0.1 Å per laser pulse. To provide additional oxygenation, the films were cooled down in $p(O_2) = 0.5$ bar.

Epitaxial SIO films were grown on different substrate materials, i. e., (110) oriented GdScO$_3$ (GSO), (110) DyScO$_3$ (DSO), (001) SrTiO$_3$ (STO), (001) (LaAlO$_3$)$_{0.3}$(Sr$_2$AlTaO$_6$)$_{0.35}$ (LSAT), and (110) NdGaO$_3$ (NGO). Table I summarizes structural parameters and surface orientation of the substrate materials. To compare structural properties of the films that were grown under different epitaxial strain conditions the film thickness was kept constant at $t \approx 60$ nm.

**TABLE I.** Space group (SG), crystallographic orientation (CO) of the surface normal, lattice parameters $a$, $b$, and $c$, orthorhombic distortion ($a/b$-1), pseudo-cubic lattice parameter $a_{pc}$ of bulk SrIrO$_3$ (SIO) [] and the used substrate materials. In addition, the corresponding lattice mismatch $\Delta = (a_{pc}(sub) - a_{pc}(SIO))/a_{pc}(SIO)$ with respect to bulk SrIrO$_3$ is noted down. For the orthorhombic substrates $a_{pc}$ is deduced from the orthorhombic lattice spacing $d_{220}$ and $d_{002}$, i. e., $a_{pc} = (d_{220}^2 + d_{002}^2)^{1/2}$. Here, LSAT is (LaAlO$_3$)$_{0.3}$(Sr$_2$AlTaO$_6$)$_{0.35}$.

|  | SG | CO | $a$ (Å) | $b$ (Å) | $c$ (Å) | $(a/b-1)(\%)$ | $a_{pc}$ (Å) | $\Delta$ (%) |
|---|---|---|---|---|---|---|---|---|
| SrIrO$_3$ | *Pbnm* |  | 5.6 | 5.57 | 7.89 | +0.52 | 3.959 | 0 |
| GdScO$_3$ | *Pbnm* | 110 | 5.48 | 5.75 | 7.93 | -4.7 | 3.969 | +0.22 |
| DyScO$_3$ | *Pbnm* | 110 | 5.44 | 5.72 | 7.91 | -4.1 | 3.950 | -0.22 |
| SrTiO$_3$ | *Pm-3m* | 001 | 3.905 | 3.905 | 3.905 | 0 | 3.905 | -1.36 |
| LSAT | *Pm-3m* | 001 | 3.874 | 3.874 | 3.874 | 0 | 3.874 | -2.14 |
| NdGaO$_3$ | *Pbnm* | 110 | 5.43 | 5.5 | 7.71 | -1.2 | 3.859 | -2.52 |

Structural properties of the SIO films, such as film thickness, surface roughness, lattice parameters and symmetry, crystallinity, epitaxial growth and strain were characterized by x-ray diffraction experiments using a Bruker D8 Diffractometer equipped with Cu$K_\alpha$ radiation ($\lambda$ = 1.5418 Å) in reflectivity, diffraction and non-coplanar grazing incidence mode.

Electronic transport was probed by four-point resistance measurements in van der Pauw geometry using a physical property measuring system (PPMS) from Quantum Design. An alternating current excitation of $I_{ac}$ = 10 µA was used. Electrical contacts for the measurements were made to the corners of the square-shaped sample surface (5 mm ×5 mm) using ultrasonic Al-wire (diameter = 15 µm) bonding. In these conditions, the so called Montgomery method [29] is particularly sensitive to anisotropic electronic transport and therefore suitable to study electronic anisotropy in, e. g., orthorhombic materials [30]. To reduce the influence of possible surface degradation on SIO film resistance [20] the measurements were carried out shortly after film deposition. In addition, the films were prepared with a rather large film thickness ($t \approx$ 60 nm) to minimize surface effects. Additional surface protection by the deposition of a thin (4 nm) epitaxial STO capping layer did not result in any changes.

III. RESULTS AND DISCUSSION

A. Structural and epitaxial properties of SrIrO$_3$ thin films

In Fig. 2 we have shown the x-ray reflectivity and diffraction (2θ/ω-scan) of epitaxial SIO films deposited with comparable thickness $t$ on various substrate materials (see Table I). Reflectivity profiles yield a similar critical angle of total reflection at $\alpha_c \approx 0.73°$ for all samples. From the best fits to the Kiessig fringes in the spectra above $\alpha_c$ we deduce a material density $D \approx 8.8$ g/cm$^3$, similar to that of bulk SIO. Furthermore, we extracted the surface roughness $R_a \approx 5$ Å and the film thickness $t \approx 60$ nm. The clear observation of Kiessig fringes and the rather low $R_a$ already indicate good layer-by-layer growth mode of SIO [31]. Structural parameters that are obtained from the fits are summarized in Table II.

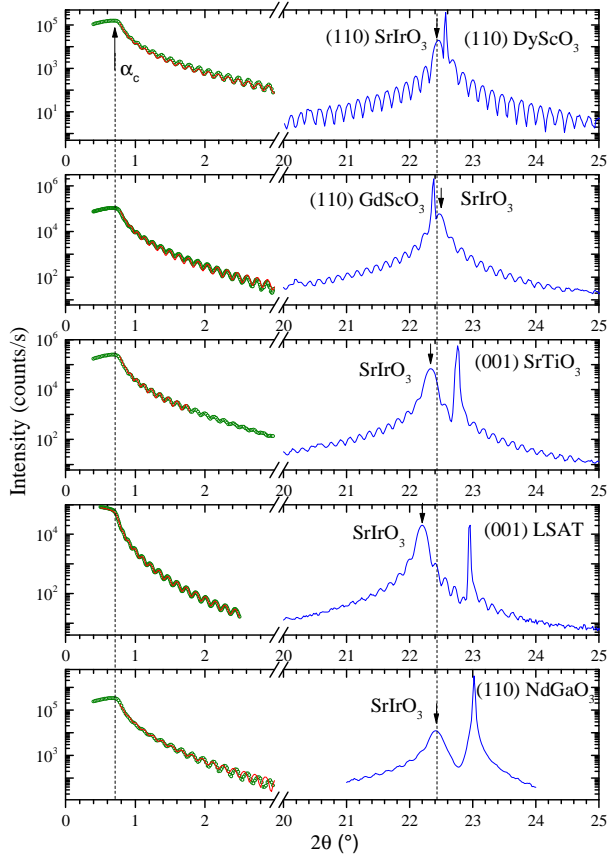

**FIG. 2.** X-ray reflectivity (left spectrum, green circles) and corresponding fit to the data (red line) of epitaxial SrIrO$_3$ thin films ($t \approx 60$ nm) deposited on various substrates. The critical angle of total reflection, $\alpha_c$, is indicated by arrow and dashed line. Symmetric x-ray diffraction ($2\theta/\omega$-scans) from lattice planes that are perpendicular to the surface normal (right spectrum). The central peak position of the film as indicated by arrow shows variation for the different substrate materials. The dashed line corresponds to the (110) peak position of bulk SrIrO$_3$.

Symmetric $2\theta/\omega$-scans of the lattice planes that are perpendicular to the surface normal are shown up to the first order reflections on the right of Fig. 2. The central peak positions of the films are seemingly sensitive to the substrate. Beside line broadening of the central peak due to the finite film thickness the diffraction displays Laue oscillations caused by coherent scattering of lattice planes, well beyond to the left and right of the central peak position, documenting an epitaxial layer-by layer growth mode and a high crystalline quality of the films. With respect to the pseudo-cubic lattice parameter $a_{pc}$ the sequence of unit-cell size is: $a_{pc}$(GSO) > $a_{pc}$(SIO) > $a_{pc}$(DSO) > $a_{pc}$(STO) > $a_{pc}$(LSAT) > $a_{pc}$(NGO), see Table I. Hence, with respect to the dashed line, indicating the (110) peak position of bulk SIO, substrate reflections are systematically shifted to the left ($a_{pc} > a_{pc}$(SIO)) or to the right ($a_{pc} < a_{pc}$(SIO)). SIO films on DSO display similar out-of-plane lattice spacing $d_{110} \approx a_{pc}$(SIO) due to the nearly perfect lattice matching $\Delta = (a_{pc}(\text{sub}) - a_{pc}(\text{SIO}))/a_{pc}(\text{SIO}) \approx -0.22$ %, where $a_{pc}$(sub) is the pseudo-cubic lattice parameter of the substrate material. Negative lattice mismatch suggests compressive in-plane strain on SIO.

With decreasing $a_{pc}$(sub), compressive strain on SIO is expected to increase. However, SIO on NGO shows bulk like $d_{110}$ lattice spacing again. The large lattice mismatch between SIO and NGO ($\Delta$ = -2.52%) very likely results in a fast structural relaxation and relieve of compressive lattice strain, indicating a small critical film thickness $t_c$ above that structural relaxation sets in.

To determine in-plane lattice parameters and orthorhombic distortion extensive x-ray diffraction measurements were carried out on various asymmetric film reflections. In Fig. 3, we have plotted 2θ/ω scans on asymmetric reflections of SIO films on DSO (top), GSO (middle), and STO (bottom) to document orthorhombicity and epitaxial relationship of SIO films. In Fig. 3 (top), the orthorhombic lattice parameters of DSO ($a < b < c$) result in the following lattice-plane spacing: $d_{260} > d_{444} > d_{620}$. As a result, the (260) reflection appears at smallest and the (620) at largest 2θ value. The (444) and (44-4) reflection have the same lattice spacing and hence 2θ position. Orthorhombic distortion, i. e., $d_{260} \neq d_{620}$ is also verified for SIO. However, here we have $b < a < c$, similar to that of bulk SIO, see Table I. Furthermore, the orthorhombic distortion is obviously much smaller for SIO (0.34%) compared to DSO (-4.1%), see Tables I and II, resulting in a much smaller difference between the 2θ-values of the (260) and (620) reflection. Nevertheless, the experimental resolution allows to exclude the presence of (444) or (44-4) reflections for the azimuth reference [-110] or [1-10]. Hence, SIO films on DSO are obviously not twinned. Compared to the orthorhombic distortion of bulk SIO (+0.52%), the distortion of the SIO film on DSO is only slightly smaller. For SIO on GSO the orthorhombic distortion is further decreased (Fig. 2, middle). Compared to the compressively strained growth of SIO on DSO ($\Delta$ = -0.22%), the growth on GSO is under tensile strain ($\Delta$ = +0.22%). Hence, tensile strain seems to reduce orthorhombic distortion, which can be well understood in terms of unit-cell size. A larger lattice parameter allows for more straightening of the Ir-O-Ir bond angle towards 180° (less buckling of $IrO_6$ octahedra) thereby reducing octahedral tilts and orthorhombic distortion in case of regular $IrO_6$ octahedra. The small orthorhombic splitting and the rather broad intensity distribution of the (260) and (444) reflections do not allow to quantify twinning of SIO. However, one may assume that the decreased orthorhombicity favors a twinned growth of SIO. For SIO on STO (Fig. 3, bottom) orthorhombic distortion seems to be negligible ($a \approx b$) or masked by twinning.

The film lattice parameters are deduced from the (110), (220), (260), (444) and (620) reflections. Table II summarizes the lattice parameters and the related orthorhombic distortion of the SIO films. Despite small changes of the orthorhombic distortion ($a/b$-1), the structural in-plane anisotropy $\sqrt{(a^2+b^2)}/c$ for (110) growth orientation is close to 1 and similar for SIO on DSO, GSO and NGO within the experimental resolution.

**TABLE II.** Structural properties of epitaxial SrIrO$_3$ (SIO) films grown on various substrates. Film thickness $t$ and surface roughness $R_a$ are deduced from x-ray reflectivity measurements. Orthorhombic lattice parameters and distortion as well as the structural in-plane anisotropy $(a^2+b^2)^{1/2}/c$ and the pseudo-cubic unit-cell volume $V_{pc} = (d_{220}^2 \times d_{002})$, where $d_{220}$ and $d_{002}$ are the orthorhombic lattice spacing, obtained from x-ray diffraction. For bulk SIO $V_{pc} \approx 61.6$ Å$^3$.

| SIO on: | $t$ (nm) | $R_a$ (Å) | $a$ (Å) | $b$ (Å) | $c$ (Å) | $(a/b-1)(\%)$ | $(a^2+b^2)^{1/2}/c$ | $V_{pc}$ (Å$^3$) |
|---|---|---|---|---|---|---|---|---|
| GSO | 58 | 5 | 5.59 | 5.58 | 7.9 | +0.17 | 0.99 | 61.6 |
| DSO | 58 | 5 | 5.61 | 5.59 | 7.92 | +0.34 | 0.99 | 62 |
| STO | 60 | 5 | 5.58 | 5.58 | 7.82 | 0 #,$ | 1.00 | 60.8 |
| LSAT | 59 | 5 | 5.6 | 5.6 | 7.82 | 0 #,$ | 1.01 | 61.3 |
| NGO | 55 | 10 | 5.59 | 5.59 | 7.91 | 0 $ | 0.99 | 61.7 |

#: SIO films on STO or LSAT display small monoclinic distortion, i. e., $\gamma = 88.82°$ and $87.54°$, respectively. $: Orthorhombic distortion might be invisible due to strong lattice relaxation or twinning.

The octahedral tilt pattern, i. e., the rotation pattern of IrO$_6$ octahedra in the SIO film is found to be the same as that of DSO and bulk SIO. The appearance of specific orthorhombic reflections, i. e., the presence of half-integer Miller indices in case of a pseudo-cubic perovskite lattice, confirms a cell doubling due to octahedral tilt or rotations. Due to the generally low intensity of such reflections, lab-source x-ray diffraction instruments are less suitable and experiments with synchrotron radiation are preferred. However, the high crystalline quality of our films allowed us to verify such reflections for the SIO films and to validate orthorhombic $a^-a^-c^+$ tilt pattern and hence *Pbnm* (62) symmetry for SIO on DSO, see APPENDIX.

SIO films that were deposited on cubic substrates with 4-fold in-plane symmetry, i. e., STO and LSAT, do not possess any sizable orthorhombic distortion ($a \approx b$), see Fig. 3 (bottom). Moreover a small monoclinic distortion appears. The monoclinic distortion ($\alpha = \beta = 90°$, $\gamma < 90°$) likely helps to compensate the lattice mismatch between the orthorhombic $a/b$-axes lattice parameter of SIO ($a_{SIO}$) and the cubic $a$-axis in-plane lattice parameter of STO or LSAT ($a_c$). We deduced the monoclinic distortion by $\sin(\gamma/2) = a_c/a_{SIO}$. Obviously, with increasing compressive lattice strain, i. e., decreasing $a_c$, monoclinic distortion increases from $\gamma = 88.82°$ for SIO on STO to $\gamma = 87.54°$ for SIO on LSAT, see Table II. However, precise refinement of the monoclinic structure was not possible because of the limited experimental resolution and number of measured reflections. Due to the four-fold in-plane symmetry of STO and LSAT, the corresponding SIO films are found to be twinned. This has been verified by the occurrence of specific (half-integer) film reflections, as detailed in the APPENDIX.

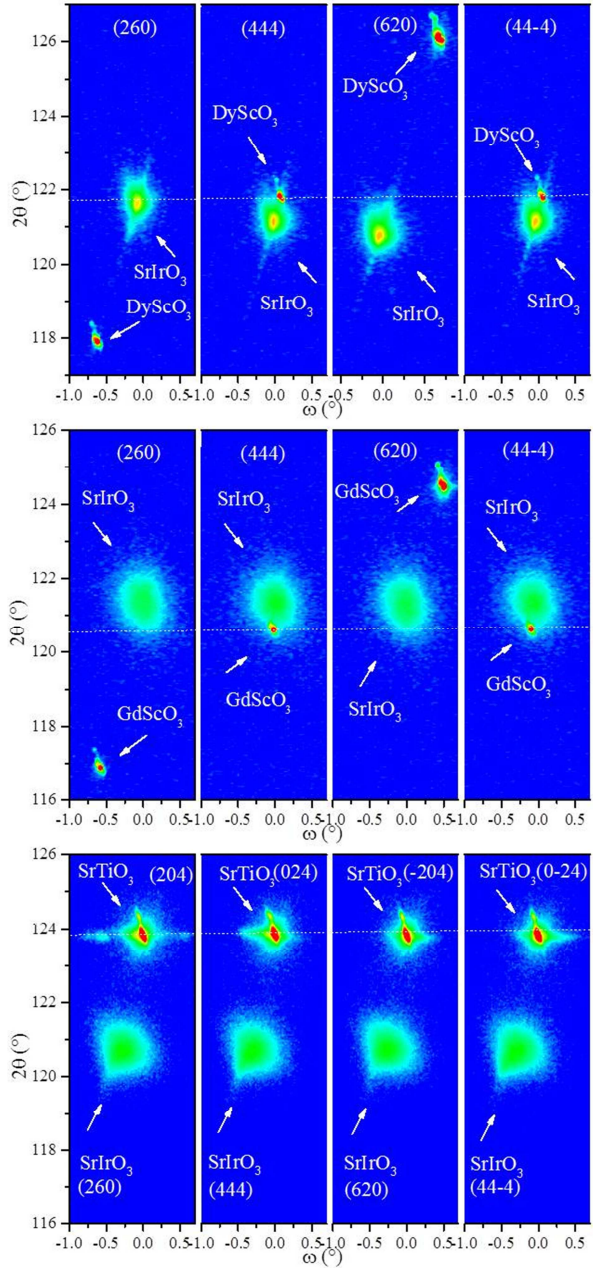

**FIG. 3.** 2θ/ω scans on asymmetric reflections of SrIrO$_3$ film on (110) DyScO$_3$ (DSO) (top), (110) GdScO$_3$ (GSO) (middle), and (001) SrTiO$_3$ (STO) (bottom). Reflections are noted with respect to orthorhombic structure. The contour plots display scattered intensity on a logarithmic scale as a function of the 2θ and ω value referring in case of DSO and GSO (STO) to the [110] ([001]) surface normal and from left to the right to the azimuth reference [-110] ([100]), [001] ([010]), [1-10] ([-100]), and [00-1] ([0-10]). For SIO reflections orthorhombic splitting decreases and intensity distribution increases from top to bottom, which hinders determination of orthorhombicity and twinning.

The epitaxial relationship of SIO with respect to the used substrate materials is sketched in Fig. 4.

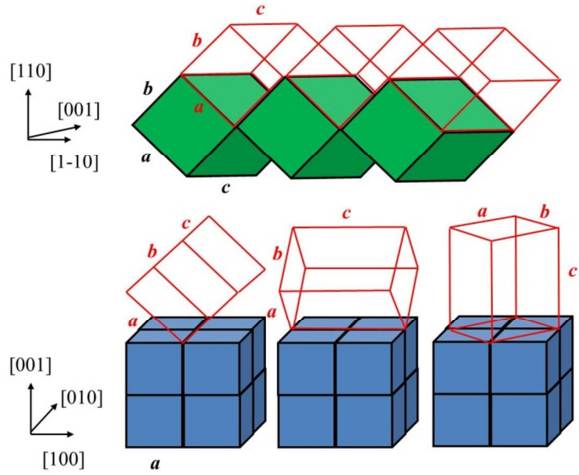

**FIG. 4.** Epitaxial relationship of SIO on DSO (top) and STO (bottom). SIO on DSO displays untwinned epitaxial growth with out-of-plane orientation: SIO[110] parallel to DSO[110] and in-plane orientation: SIO[001] parallel to DSO[001]. In contrast, twinned growth is observed for SIO on STO due to the four-fold in-plane symmetry of the substrate material. Besides in-plane twinning (left and middle) also out-of-plane twinning (middle and right) may occur. Substrate orientation and axes are indicated in black. SIO lattice and axes are given in red.

Since the SIO films were grown on substrate materials displaying different lattice mismatch with respect to the crystal structure of bulk SIO, the films are likely epitaxially strained. Generally, lattice strain caused by lattice mismatch between film and substrate material is relieved and compensated during film growth by introducing defects or lattice distortions above a so-called critical film thickness $t_{cr}$ resulting in a change of the lattice parameters with increasing film thickness for $t > t_{cr}$. In Table III we have summarized the lattice strain of SIO films grown on the various substrate materials. Strain was deduced from the lattice parameters given in Tables I and II, and therefore has to be regarded rather as a "mean" lattice strain. The strain values listed in Table III are rather small and do not show any systematic trend with decreasing in-plane lattice parameters of the substrate, i. e., from GSO to NGO (from top to bottom of Table III). This indicates a nearly strain-relieved, bulk-like state of the films. Irrespective of the substrate, the *b*-axis seems to be under a small tensile strain. A sizeable compressive strain is only found for the *c*-axis for the films grown on STO or LSAT. Since *a*- and *b*-axis components are present in in-plane as well as in out-of-plane direction, frustration with respect to lattice strain may occur. That might be the reason for the relative small value and change of $\Delta d_{220}$ with increasing compressive lattice mismatch. Apparently, the most part of lattice strain is comprised within the *c*-axis lattice parameter. Further studies on the epitaxial strain and related strain-relaxation are detailed in the APPENDIX.

In summary, SIO films on DSO substrates have *Pbnm* symmetry with the octahedral tilt pattern of $a^-a^-c^+$ as observed in bulk SIO. SIO thin films on DSO are not twinned, whereas epitaxial growth on cubic substrate materials results in a twinning of SIO. Due to the limited experimental resolution in lab-based x-ray diffractometers and the limited number of accessible film reflections we cannot precisely determine symmetry group, orthorhombicity, or the degree of twinning of SIO grown on GSO, STO, LSAT or NGO. However, one may assume that the possibility of twinning increases with decreasing orthorhombic distortion of the substrate material, i. e., smallest for SIO on DSO, medium for SIO on GSO and largest for SIO on STO. Increasing lattice mismatch may further favor a twinned growth for energetic reasons. Generally, film strain is found to be nearly relaxed or rather small well in consistence with the observed bulk-like structural properties of the SIO films. Our findings are in good agreement with literature [32,33].

**TABLE III.** Epitaxial strain of SrIrO$_3$ (SIO) films grown on various substrates with respect to bulk SIO. Film thickness $t$ was kept nearly constant for all the films ($t \approx 60$ nm). The "mean" strain $\Delta a = (a_f - a_b)/a_b$, $\Delta b = (b_f - b_b)/b_b$, $\Delta c = (c_f - c_b)/c_b$, $\Delta d_{220} = (d_{220f} - d_{220b})/d_{220b}$, and $\Delta V_{pc} = (V_{pcf} - V_{pcb})/V_{pcb}$ was calculated from the "mean" structural parameters of the SIO films $a_f$, $b_f$, $c_f$, $d_{220f} = 1/2 \times (a_f^2 + b_f^2)^{1/2}$, and $V_{pcf}$ (see Table II) and bulk SIO $a_b$, $b_b$, $c_b$, $d_{220b} = 1/2 \times (a_b^2 + b_b^2)^{1/2}$, and $V_{pcb}$ (see Table I). Positive or negative signs indicate tensile or compressive strain, respectively. The pseudo-cubic unit-cell volume $V_{pc} = (d_{220}^2 \times d_{002})$, where $d_{220}$ and $d_{002}$ are the orthorhombic lattice spacing. For bulk SIO $V_{pc} \approx 61.6$ Å$^3$. Films on STO or LSAT experience a monoclinic distortion, i. e., $\gamma < 90°$, which has been neglected in the calculation of $V_{pc}$.

| SIO on: | $\Delta a$ (%) | $\Delta b$ (%) | $\Delta c$ (%) | $\Delta d_{220}$ (%) | $\Delta V_{pc}$ (%) |
|---|---|---|---|---|---|
| GSO | -0.17 | +0.17 | +0.12 | 0 | 0 |
| DSO | +0.17 | +0.35 | +0.38 | +0.25 | +0.64 |
| STO | -0.35 | +0.17 | -0.88 | 0 | -1.29 |
| LSAT | 0 | +0.53 | -0.88 | +0.25 | -0.48 |
| NGO | -0.17 | +0.35 | +0.25 | +0.12 | +0.16 |

B. Electronic transport

For the resistance measurements shown in Fig. 5, the films were first cooled down from $T = 300$ K to $T = 2$ K and then heated up again to 300 K. Heating and cooling rate was kept constant at about 1K/min to ensure good thermalization of the sample. In the figure we have displayed the resistance versus temperature of SIO on GSO, DSO, STO, and NGO for the two orthogonal directions along the substrate edges. Strictly speaking, determination of the resistivity from van der Pauw measurements is only valid for systems displaying a homogeneous conductivity. For that reason we preferred to give here only the measured resistance. As illustrated in Fig. 6 a), the specific resistances on the orthorhombic (cubic) substrates are obtained as follows: $R_{1-10}$ ($R_{010}$) = $U_{21}/I_{34}$ and $R_{001}$ ($R_{100}$) = $U_{41}/I_{32}$. Note that cooling and heating curves match perfectly with each other. Interestingly, a distinct difference between $R_{1-10}$ and $R_{001}$ up to 25% is observed for SIO on GSO and DSO. This difference cannot be explained by considering only geometric factors, e. g., different spacing between the electrical contacts or even inhomogeneity in film thicknesses. Therefore, intrinsic anisotropic electronic transport is more obvious. The larger resistance is observed along the [1-10] direction. Moreover, for 300 K > $T$ > 200 K, $R_{1-10}$ is nearly constant or increases only slightly with decreasing $T$, whereas $R_{001}$ decreases and passes a shallow minimum around 200 K. Below 200 K, both resistances show a distinct increase with decreasing $T$. Nevertheless, the resistance ratio $R(2\ K)/R(300\ K)$, which is obviously somewhat larger for SIO on GSO and DSO compared to SIO on STO or NGO, is rather small ($\leq 2$). The resistivity of the SIO film on STO at 2 K is rather large and amounts to $\rho \approx 1$ mΩcm which indicates, with respect to the Ioffe-Regel criterion [34], electronic transport close to a MIT. The absolute value of $\rho$ is in good agreement with the ones of bulk SIO and SIO films ($t$ > 3 nm) that are reported in literature [19,23,24].

For SIO on NGO the resistance anisotropy is much smaller and for the twinned SIO films on STO or LSAT (not shown) even negligible. Here, the $R$ versus $T$ behavior is very similar to that of $R_{001}$ observed for SIO on GSO or DSO. These observations strongly suggest that anisotropic resistance is caused by the orthorhombic distortion of the SIO films.

Band-structure calculations on SIO [11,23] indeed display high sensitivity towards orthorhombic distortion and lattice strain.

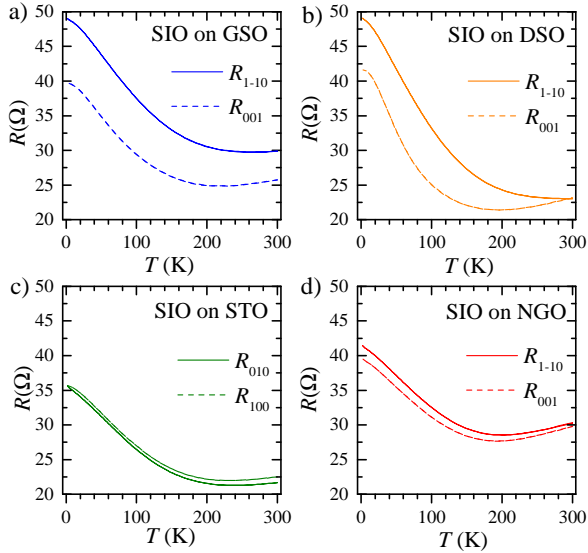

**FIG. 5.** Resistances $R_{1-10}$ and $R_{001}$ for the two orthogonal in-plane directions of the substrate versus $T$ for SIO on GSO (a), DSO (b), and NGO (d). For definition of the resistances, see Fig. 6. (c) Corresponding resistances $R_{010}$ and $R_{100}$ for SIO on STO. The measurements were carried out during thermal cycling from 300 K to 2 K and from 2 K to 300 K. Heating and cooling rate was kept constant at about 1K/min to ensure good thermalization of the sample. Note, that cooling and heating curves match each other perfectly.

Fig. 6 b) shows the Hall resistance $R_H = U_{42}/I_{13}$ for SIO on STO at 2 K and 300 K. The linear $B$-dependence and negative slope of $R_H(B)$ indicate a dominant single-band electron-like transport over the complete $T$-range. As demonstrated by ARPES measurements [11], SIO displays a semi-metallic behavior, i. e, electron-like and hole-like pockets close to the Fermi surface. However, effective mass of holes was found to be about 2-6 times larger than the one of electrons, and hence they are less mobile. Moreover, a downshift of hole-like bands is expected for SIO on STO [23], which may justify the assumption of a single electron band to determine charge carrier density. From Fig. 6 b) the deduced electron concentration is $n_e = 9.5 \times 10^{20}$ cm$^{-3}$ at 300 K and $n_e = 1.4 \times 10^{20}$ cm$^{-3}$ at 2 K. The values of $n_e$ agree well with the ones obtained from ARPES measurements ($4.7 \times 10^{20}$ cm$^{-3}$) [11]. The decrease of $n_e$ with decreasing $T$ is likely due to charge trapping by structural defects, which seems to be typical for oxides [35]. The electron mobility increases from $\mu_e = 12$ cm$^2$/(Vs) at 300 K to 47 cm$^2$/(Vs) at 2 K.

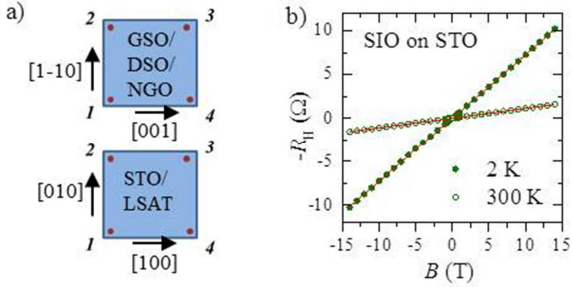

**FIG. 6.** (a) Schematic of the measurement set-up. Four contacts are bonded to the corners of the 5 mm × 5 mm square-shaped sample surface. Substrate edges are parallel to the [1-10] and [001] directions for GSO, DSO, and NGO and parallel to the [010] and [100] directions for STO and LSAT. The resistances are measured in four-point van der Pauw geometry and are defined as follows: $R_{1\text{-}10}$ and $R_{010} = U_{21}/I_{34}$; $R_{001}$ and $R_{100} = U_{41}/I_{32}$; and $R_H = U_{42}/I_{13}$. (b) Negative Hall resistance $R_H$ (symbols) versus magnetic field $B$ for SIO on STO at $T = 2$ K and 300 K. The solid lines are linear fits to the data.

To analyze the temperature dependence of the anisotropic resistance in more detail, we have plotted the normalized resistance ratio $r_N$ as function of $T$ in Fig. 7. To this end, the resistances $R_{1\text{-}10}(T)$, $R_{001}(T)$, $R_{010}(T)$, and $R_{100}(T)$ were first normalized to its room temperature values, $r = R(T)/R(300\text{K})$, and then the ratio $r_N$ between the two orthogonal resistances $r_{1\text{-}10}/r_{001}$ and $r_{010}/r_{100}$ were calculated. As already visible in Fig. 5, the anisotropic behavior is well pronounced for SIO on DSO, moderate for SIO on GSO, and very small for SIO on STO and NGO and absent for SIO on LSAT ($r_N(T) \approx 1$). Interestingly, $r_N$ displays a distinct $T$-dependence. For SIO on DSO or GSO $r_N$ first increases with decreasing $T$, peaks around 86 K and then decreases again.

The distinct $T$-dependence of $r_N$ is very similar to that of the structural in-plane anisotropy with respect to the [1-10] and [001] direction of bulk SIO. The structural in-plane anisotropy $\sqrt{(a^2+b^2)}/c$ of bulk SIO is shown versus $T$ in Fig. 7 b). Data were taken from the work of Blanchard et al. (Ref. [13]). With respect to the nearly strain relived state of all the SIO films and the more or less bulk-like lattice parameters, a comparison to bulk lattice parameters can be justified. Note, that the structural in-plane anisotropies of the SIO films are comparable at room temperature, well consistent with a strain-relieved structure. For SIO films, $\sqrt{(a^2+b^2)}/c$ differs only slightly from the bulk value (cf. Table II). The small "mean" strain on the $c$-axis (see Table III) is likely responsible for that. For bulk SIO, the $c$-lattice parameter decreases stronger compared to $\sqrt{(a^2+b^2)}$ with decreasing $T$ down to $T_{max} \approx 85$ K, where the anisotropy displays a clear maximum. Below $T_{max}$ the lattice parameters are nearly constant except $a$ which still further decreases slightly with decreasing $T$ and reduces the anisotropy again (see Ref. [13]). The decrease of the unit cell parameters in SIO is found to be solely caused by changes of bond-lengths. Tilt angles of $IrO_6$ octahedra are essentially independent of $T$ below 300 K [13]. Therefore, with decreasing $T$ down to about 85 K, hybridization of oxygen $2p$ and Ir $5d$ orbitals is more enhanced along the [001]- than the [110] direction of SIO, which is very likely the reason for the increase of $r_N = r_{1-10}/r_{001}$. Note, with respect to the small structural changes versus $T$, changes of $r_N$ are two orders of magnitude larger. Reason for that high sensitivity of the electronic transport towards structural changes could be probably due to the narrow bandwidth of the electronic band structure. The octahedral distortion in combination with the rather large SOC in SIO results in multiple bands depending very sensitively on octahedral tilt and lattice strain [23]. For example, ARPES measurements by Schütz and coworkers [25] revealed that the thickness-dependent metal-insulator transition in SIO films on STO at $t < 3$ unit cells [24] is possibly caused by a suppression of in-plane octahedral rotations which seems to be imposed by the STO substrate. This apparently underscores the high sensitivity of electronic transport with respect to orthorhombic distortions. Therefore, substrate induced structural changes of the SIO film may indeed be relevant for pseudomorphic films but should not be significant for rather thick and strain-relaxed films as used here. Nevertheless, small hysteretic features of $r_N$ between heating and cooling visible for SIO on STO and NGO might indicate substrate induced electronic anisotropy in our SIO films alike. For SIO on STO and NGO, the $T$-dependence of $r_N$ displays a clear hysteretic behavior during thermal cycling. To underline such hysteretic character we have shown $r_N$ on an enlarged scale in Fig. 7 c). For SIO on STO $r_N(T)$ becomes different below $T = 105$ K for the cooling and heating curves. The resistance hysteresis persists down to 60 K. Below about 60 K, cooling and heating curves coincide again. Note, that heating and cooling rate were kept constant at about 1K/min to ensure good thermalization of the sample. Furthermore, that feature was also observed for many other SIO films prepared on STO and hence well reproducible. Remarkably, hysteretic behavior occurs at similar $T$ where STO shows an antiferrodistortive (AFD) cubic-to-tetragonal phase transition [36]. This AFD transition results in an anti-phase rotation of the $TiO_6$ octahedra along the $c$-axis, i. e., $a^0a^0c^-$ according to Glazer`s notation. The exact mechanism involving the degree of rotational correlations between $IrO_6$ and $TiO_6$ octahedra is certainly complicated and currently beyond the scope of this work. In addition, relaxation of structural mismatch with increasing film thickness may hinder a quantitative analysis.

For SIO on NGO similar hysteretic behavior is observed above 270 K. Interestingly, at 240 K NGO displays structural anomalies alike, which are related to an isosymmetric transition that results in an anomalous hysteretic thermal expansion of the lattice parameters [37-39]. Therefore, the characteristic features at 105 K and 270 K for SIO on STO and NGO, respectively, could be very likely related to structural changes in the substrate material. It is not unusual that the octahedral tilt pattern of perovskite films is affected by the substrate material [40]. However, it is remarkable, that electronic transport in SIO is sensitive to such presumably small structural changes. On the other side, small deviations between heating and cooling curves are also present for SIO on DSO and GSO. Unfortunately, to the best of our knowledge there are no data on the thermal expansion of these materials available which might help to clarify these features. Therefore, the appearance of those small hysteretic features is still rather unclear and has to be investigated in more detail in future work.

Since the structural in-plane anisotropies of the strain-relieved SIO films are similar, the steady decrease of the anisotropy of $r_N$ for SIO on GSO to SIO on NGO, STO, and LSAT is most likely caused by an increased twinned film growth which is favored by a square-shaped (four-fold) substrate surface-cell or by an increase of the in-plane lattice-mismatch between SIO and substrate material.

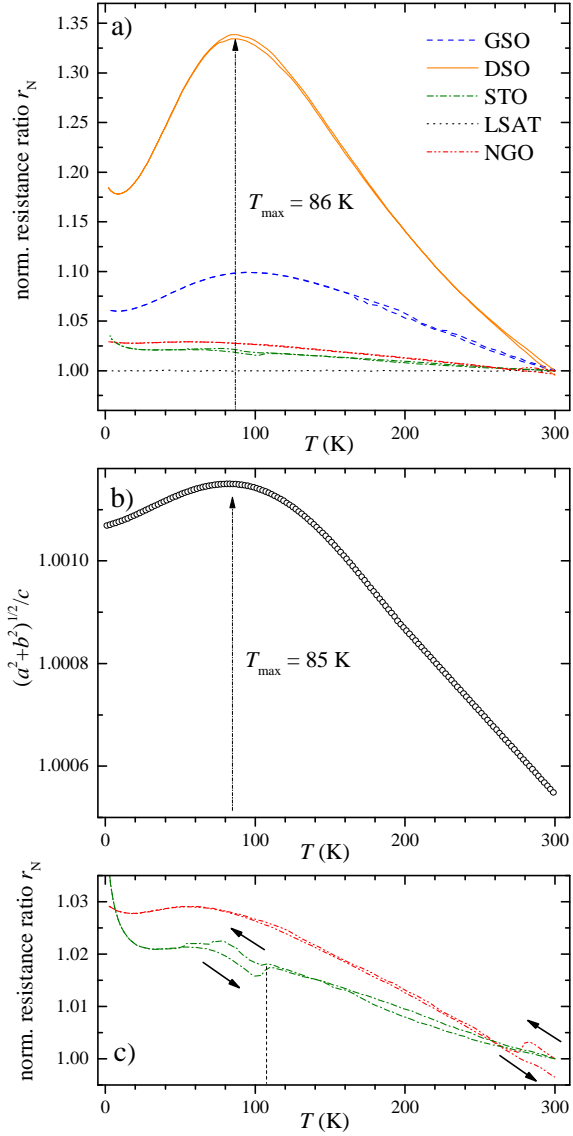

**FIG. 7.** (a) Normalized resistance ratio $r_N$, see text, versus $T$ for SIO on DSO, GSO, NGO, STO and LSAT (from top to bottom). First, the samples were cooled down from 300 K to 2 K and then heated up again to 300 K. Heating and cooling rate was kept constant at about 1K/min to ensure good thermalization of the sample. (b) $T$-dependence of the structural anisotropy $\sqrt{(a^2+b^2)}/c$ of bulk SIO. Lattice parameters were taken from Ref. [13]. (c) Enlarged scale of $r_N$ versus $T$ for SIO on STO and NGO. Sample cooling and heating curves are indicated by arrows displaying differences in the range of 70 K - 105 K for SIO on STO and above 270 K for SIO on NGO. The antiferrodistortive transition of STO at 105 K is indicated by the dashed line.

## IV. SUMMARY

Perovskite SrIrO$_3$ (SIO) films were grown epitaxially on various substrate materials by pulsed laser deposition. Films grown on orthorhombic (110) DSO display untwinned bulk-like orthorhombic structure with space group *Pbnm*. The lattice parameters and the structural in-plane anisotropies for the SIO films are similar, indicating a strain-relived state of SIO. Films deposited on cubic STO show twinned growth. Twinning likely increases steadily for SIO on GSO, NGO, STO, and LSAT favored by a square-shaped (four-fold) substrate surface-cell or by an increase of the in-plane lattice-mismatch between SIO and substrate material. Generally, lattice strain due to the epitaxial growth seems to be relaxed to large extent for the rather thick (60 nm) SIO films. Resistance measurements on the SIO films reveal distinct anisotropic behavior. For SIO on DSO, resistance $R_{001}$ along the [001] direction of SIO is found to be smaller compared to the resistance $R_{1-10}$ along the [1-10] direction. The film resistance for the two orthogonal directions differs by up to 25% indicating distinct anisotropic behavior. The resistivity of $\rho \approx 1$ mΩcm at 2 K is well comparable to that of bulk SIO. Hall measurements indicate dominant electron-like transport throughout the temperature range from 2 K – 300 K. The anisotropy of the resistance $r_N$ for untwinned SIO on DSO shows a pronounced *T*-dependence with a maximum at 86 K. However, for twinned SIO films grown on STO anisotropic behavior nearly vanishes. The distinct *T*-dependence of $r_N$ is similar to that of the structural in-plane anisotropy $\sqrt{(a^2+b^2)}/c$ caused by the orthorhombic distortion of SIO. Therefore, the anisotropic electronic transport is very likely related to the orthorhombic distortion of SIO. The disappearance of anisotropy for twinned films strongly supports that assumption. Small hysteretic behavior of $r_N$ at $T = 105$ K and 270 K found for SIO on STO and NGO, respectively, indicate that structural changes of the substrate material affect electronic transport and anisotropy alike. The substrate induced effects are likely related to constraints with respect to $IrO_6$ octahedral rotations. The strong sensitivity of the electronic transport in SIO films to even small structural changes may be explained in terms of the narrow electron-like bands in SIO caused by SOC and orthorhombic distortion.

ACKNOWLEDGEMENTS

We are grateful to R. Thelen and the Karlsruhe Nano Micro Facility (KNMF) for technical support. DF also acknowledges A. Beck and A. Zaitsev for SEM analysis and K. Sen for fruitful discussions. Thanks to K. Grube for providing digitized data.

V. APPENDIX

A. Orthorhombic distortion in SIO films

With reference to a *doubled pseudo-cubic perovskite cell*, octahedral antiphase rotations of ($a^-$) or ($b^-$) produce reflections with $k \neq l$, or $h \neq l$, respectively (*h*, *k*, and *l* are odd) e. g., (131), (113), and (311), whereas in-phase rotations ($c^+$) produces reflections with $h \neq k$ (*h* and *k* are odd, and *l* is even), e. g., (130) or (310) [41]. Due to the generally low intensity of such reflections, lab-based x-ray diffraction instruments are less suitable for these studies and experiments with synchrotron radiation are preferred. However, the high crystalline quality of the films allowed us to successfully verify such reflections for the SIO films. Figure 8 shows 2θ/ω scans on such pseudo-cubic (*single pseudo-cubic perovskite cell*) half-integer reflections of SIO on DSO. The pseudo-cubic $a^*$, $b^*$, and $c^*$-axis were chosen parallel to the orthorhombic [001], [1-10], and [110] directions, respectively. With respect to the orthorhombic symmetry group and octahedral tilt pattern, only specific half-integer reflections are observed. In contrast, the corresponding pseudo-cubic integer film and substrate reflections are all symmetry-allowed and present (not shown here). The (130) and (310) reflections are verified by measurements under grazing incidence (GID) which gives confidence on the absence of the half-integer (1/2 3/2 0) and (3/2 ½ 0) reflections. The observed half-integer reflections document $a^+b^-c^-$ tilt-pattern for both SIO and DSO, consistent with *Pbnm* (62) symmetry and $a^-b^-c^+$ if $a^*$, $b^*$, and $c^*$-axis were chosen parallel to the orthorhombic [110], [1-10], and [001] directions, respectively. Moreover, the absence of twinning is documented alike, consistent with the measurements shown in Fig. 3.

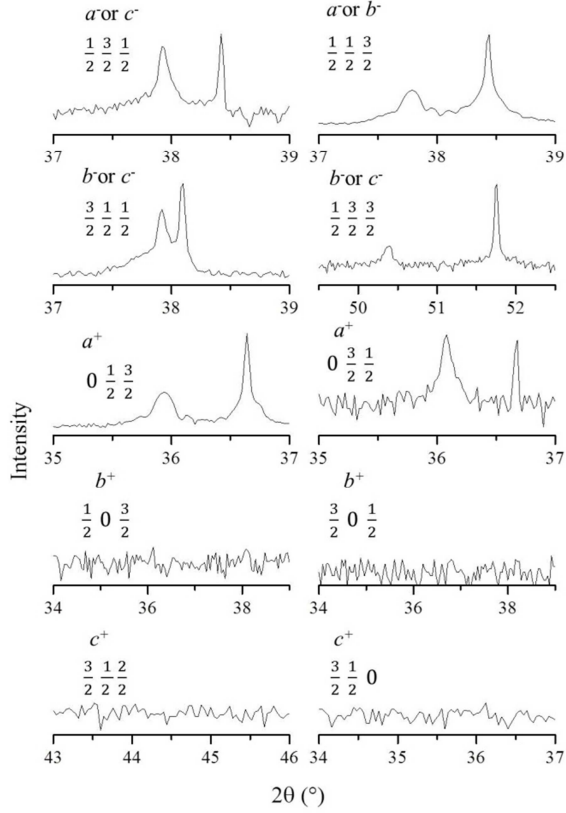

**FIG. 8.** 2θ/ω scans on pseudo-cubic half-integer reflections of SrIrO$_3$ film on DyScO$_3$ substrate. The pseudo-cubic $a^*$, $b^*$, and $c^*$-axis were chosen parallel to the orthorhombic [001], [1-10], and [110] directions, respectively. The intensity is given on a logarithmic scale. The measurements were carried out in symmetric diffraction conditions. With respect to the octahedral tilt pattern, i. e., in-phase rotation (+) or out-of-phase rotation (-) around $a^*$, $b^*$, or $c^*$-rotation axis - only specific half-integer reflections (as indicated) should be observed. In contrast, the corresponding pseudo-cubic integer film and substrate reflections are all symmetry-allowed and observed (not shown here). The 310 and $^3/_2\,^1/_2\,0$ were measured in grazing incidence geometry. The film reflection appears always to the left side of the substrate reflection. Observed half-integer reflections document $a^+b^-c^-$ tilt-pattern for both SIO and DSO, consistent with *Pbnm* (62) symmetry.

The epitaxial growth of SIO on cubic substrates such as STO or LSAT is expected to result in twinned films because of the four-fold in-plane symmetry. We likewise carried out 2θ/ω scans on pseudo-cubic half-integer reflections of SrIrO$_3$ to verify twinned growth of SIO on these substrates. In Fig. 9, we have shown half-integer reflections of SIO with respect to the pseudo-cubic perovskite cell (see above). The observed half-integer film reflections are only consistent with the presence of two domains displaying an $a^+b^-c^-$ and $b^-a^+c^-$ tilt-pattern, respectively, which means an in-plane twinning of SIO. The individual tilt pattern is the same as that of bulk SrIrO$_3$.

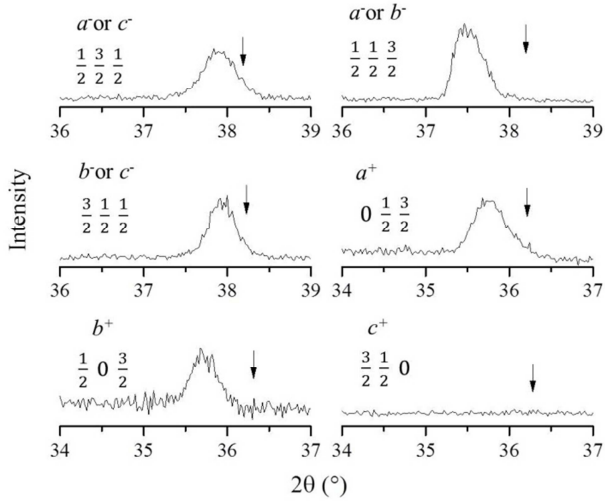

**FIG. 9.** 2θ/ω scans on pseudo-cubic half-integer reflections of $SrIrO_3$ film on $SrTiO_3$ substrate. The pseudo-cubic $a^*$, $b^*$, and $c^*$-axis were chosen parallel to the orthorhombic [001], [1-10], and [110] directions, respectively. The intensity is given on a linear scale. Maximum peak intensity is well comparable between all the reflections. In contrast to the integer pseudo cubic reflections, which are all observed for both, film and substrate (not shown here), half-integer reflections are only visible for $SrIrO_3$ and not for cubic $SrTiO_3$, which would appear, as indicated by the arrow, to the right of the film reflection. With respect to the orthorhombic symmetry group and octahedral tilt pattern, i. e., in-phase rotation (+) or out-of-phase rotation (-) around $a^*$, $b^*$, or $c^*$-rotation axis - only specific half-integer reflections are observed. Observed half-integer reflections are only consistent with the presence of two domains displaying an $a^+b^-c^-$ and $b^-a^+c^-$ tilt-pattern, i. e., an in-plane twinning of $SrIrO_3$. The individual tilt pattern is the same as that of bulk $SrIrO_3$.

### B. Epitaxial strain in SIO films

To study epitaxial strain and lattice relaxation in more detail, we carried out reciprocal lattice mapping on asymmetric reflections, which allows us to analyze intensity distribution along the in- and out-of-plane direction of the substrate materials. In Fig. 9 we have shown exemplarily reciprocal space maps of the SIO films grown on DSO, STO, and NGO. Since the strain-state of the SIO films does not differ so much (see Table III), an increasing strain relaxation is expected with increasing lattice mismatch Δ if a pseudomorphic growth is assumed at the beginning of the growth process. That trend is indeed observed in Fig. 10. Small lattice compression of SIO on DSO results in rather negligible lattice relaxation. Therefore, the (332) film peak displays symmetric and sharp intensity distribution with respect to the in-plane direction. For SIO on STO the lattice spacing difference Δ is larger, hence leading to a stronger smearing-out of the (332) SIO peak intensity towards the peak position of bulk SIO. For SIO on NGO the lattice mismatch is largest amounting to Δ = -2.52%. Such a high lattice mismatch usually generates very rapid lattice relaxation, i. e., $t_{cr}$ is very small. Therefore, most part of the scattered intensity originates from nearly fully relaxed film material leading again to a rather symmetric intensity distribution. The strong lattice relaxation is also very likely the reason for the increased surface roughness of SIO on NGO, see Table II. Strain relaxation usually increases the mosaic spread of the crystal structure alike, resulting in an additional decrease of peak intensity. This trend can also be well observed in Fig. 10, where the maximum peak intensity of the (332) SIO reflection clearly decreases from left (SIO on DSO) to the right (SIO on NGO).

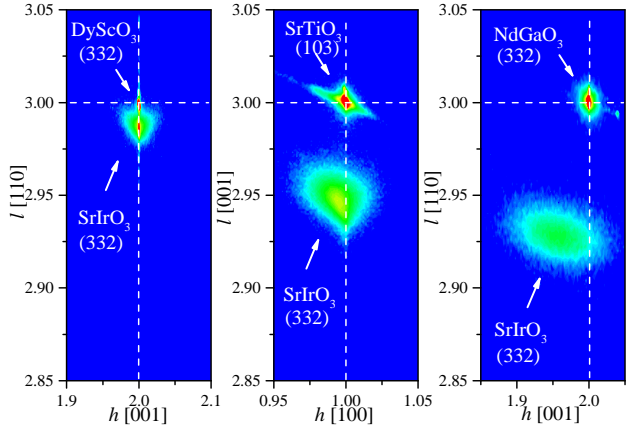

**FIG. 10.** Contour plots displaying reciprocal space maps of SrIrO$_3$ thin films grown on (110) DyScO$_3$ (left), (001) SrTiO$_3$ (middle), and (110) NdGaO$_3$ (right). The maps are recorded in the vicinity of the (332) SrIrO$_3$ reflection. The scattered intensity is given on a logarithmic scale and plotted as a function of the scattering vector $q$ expressed in noninteger Miller indices $h,k$, and $l$ of the substrate reflection, referring to the azimuth reference [001] ([100]) and the surface normal [110] ([001]) for DyScO$_3$ and NdGaO$_3$ (SrTiO$_3$). Lattice reflections are indicated. In-plane and out-of-plane reciprocal lattice spacing of the substrate reflection are marked by dashed lines.